\documentclass[preprint,12pt]{elsarticle}
\usepackage{natbib} 
\usepackage{mathtools}
\usepackage{amssymb}
\usepackage{circuitikz}
\usepackage{graphicx}        
\usepackage{caption}  
\usepackage{subcaption} 
\usepackage{booktabs}
\usepackage{tabularx}
\usepackage{multirow}

\makeatother
\journal{ }
\begin{document}
	
	\begin{frontmatter}
		\title{\textit{ten}SVD algorithm for compression}
		
		\author{Michele Gallo}
		\affiliation{organization={University of Naples - L'Orientale},
			Department of Human and Social Sciences,\\
			addressline={L.go S.Giovanni Maggiore, 30},
			city={Naples},
			postcode={80134},\\
			country={Italy}, \\
			email={mgallo@unior.it}}

\begin{abstract}
		Tensors provide a robust framework for managing high-dimensional data. Consequently, tensor analysis has emerged as an active research area in various domains, including machine learning, signal processing, computer vision, graph analysis, and data mining. This study introduces an efficient image storage approach utilizing tensors, aiming to minimize memory to store, bandwidth to transmit and energy to processing. The proposed method organizes original data into a higher-order tensor and applies the Tucker model for compression. Implemented in R, this method is compared to a baseline algorithm. The evaluation focuses on efficient of the algorithm measured in term of computational time and the quality of information preserved, using both simulated and real datasets. A detailed analysis of the results is conducted, employing established quantitative metrics, with significant attention paid to sustainability in terms of energy consumption across algorithms. 
\end{abstract}

	\end{frontmatter}

\section{Introduction}
\label{sec:intro}
A large volume and complexity of image and video datasets are generated by modern devices every second. All this data necessitate continuous improvement of methods to reduce storage space and processing time, thereby facilitating more efficient extraction and transmission of information. It is well known that there are two major categories of methods for reduce memory space. The latter guarantees substantial reductions in memory cost while maintaining a high percentage of original information. Ongoing research endeavors persist in exploring and developing innovative methods to address the challenges associated with compression tasks.\\
To this end, a diverse array of methodologies has been proposed across various disciplines, including astronomy, biology, chemistry, engineering, geology, and medical imaging. Several comprehensive surveys have been conducted to organize the different approaches employed in these domains, with \citep{JAMIL2023106361} and \citep{capitan2015recent} being notable recent surveys comparing several techniques for image and video compression.\\
It is well established that image compression can be effectively achieved utilizing multivariate techniques such as Singular Value Decomposition (SVD) \citep{Stewart93}, \citep{Nayar96}. While grayscale images are typically represented as two-dimensional matrices, color images are commonly structured in three dimensions (RGB channels). These multivariate methods can be applied to diminish the dimensionality of each color channel independently. However, such approaches become inefficient when dealing with videos or more complex signals that necessitate consideration of higher-dimensional data.\\	
Consequently, methodologies such as Tucker analysis \citep{Tucker:1966} and broader tensor decompositions are increasingly leveraged for compression purposes. For an overview of tensor decomposition models, refer to \citep{NIE2023667}. These approaches are better equipped to manage the complexity inherent in high quality of images or videos  and their convert to higher-dimensional data, thereby enhancing both compression efficiency and overall performance.	
A significant bottleneck for all compression methods is the need to achieve only the helpful information included in an high resolution images or videos. In this way a low storage footprint is  need while maintaining an acceptable level of quality. The trade-off between reduced storage and the details and clarity of images or videos is consistently considered in compression contexts. Additionally, the time required for compression processes is a critical factor in signal transmission, as it directly influences transmission efficiency and bandwidth utilization. Rarely compression is studied as a fundamental tool for achieving in  a more sustainable way images and videos. However, concerns regarding the energy consumption of data centers are rapidly increasing, and more of the huge amounts of energy that their consumption is for processing and archiving the million of images and videos that are collected every day. Thus, we must pay more attention when developing algorithms for image and video compression.\\
Based on tensor decomposition, a novel algorithm is proposed for this aim, called \textit{ten}SVD (tensor Singular Value Decomposition). It is designed to ensure high sustainability in terms of energy consumption but balancing the trade-off between storage and quality requirements.\\
To ensure a fair comparison, processing time, particularly compression time with the same CPU, serves as an effective proxy for evaluation efficiency.  Thus, \textit{ten}SVD is compared to the established algorithm for Tucker analysis, Higher-Order Singular Value Decomposition (HOSVD) \citep{de2000best}. The strengths and weaknesses of \textit{ten}SVD are analyzed in terms of several metrics, including Peak Signal-to-Noise Ratio (PSNR), Mean Squared Error (MSE), and Processing Time (TIME).\\
The remainder of this article is structured as follows. Section~\ref{sec:2} introduces tensor, including preliminary concept. Section~\ref{sec:3} presents the main idea of approach proposed and relative algorithm. Case study with full comparative results are in Sections~\ref{sec:4}. Finally, in Section~\ref{sec:5} are given conclusions and findings.

		\section{Tensor decomposition}
\label{sec:2}
Tucker model (Tucker3) is one of the most cited and applied methods for tensor decomposition. Originally proposed for three-way cases, it is referred to in several ways, including Three-Mode Principal Component Analysis (3MPCA), N-Mode Components Analysis (N-Mode PCA), and Higher-Order SVD (HOSVD) \citep{Tucker:1966, kroonenberg1980, kapteyn1986approach, de2000multilinear}. Today, the Tucker model has been generalized to higher-order tensors and it is quickly increasing its application for imaging analysis \citep{bip2021tensors}. For clarity, some preliminary concepts and notations will be recalled in Section~\ref{sec:2.1} and will be used throughout the paper.

\subsection{Preliminary concepts and notations}
\label{sec:2.1}
Objects with more than two indices are referred to as multi-way arrays, whereby the number of indices is termed the order. Nowadays, there is a trend to call any multi-way array as a tensor, and we keep this tradition to identify the linear operators with their multidimensional objects  \citep{trendafilov:mda}. \\
A real-valued tensor $\mathcal{X} \in \mathbb{R}$ of order $N \in \mathbb{N}$ is denoted by indices $(I_1 \times \dots \times I_N)$, and its entries by $x_{i_1  \dots i_N}$. 
When only a fixed subset of indices is considered, this part of tensor is termed a \textit{subtensor}. The fibers are defined by fixing every index except one, i.e. $\mathbf{x}(:,i_2, i_3, \dots, i_N)$,  and slices obtained by fixing all but two indices, i.e. $\mathbf{X}(:,:,i_3,\dots,i_N)$. \\
Tensors can be reformatting in several ways, and a particular case of manipulation is called \textit{n-}mode unfolding or matrix unfolding. Denoted by $\mathbf{X}_{(n)}$, the \textit{n}-mode  unfolding of a tensor $\mathcal{X}$ arranges the \textit{n}-mode fibers as columns in the resulting matrix:  $\mathbf{X}_{(n)}\ (I_n \times \prod_{i\neq n}I_i)$. \\ 
According to the Kolda and Bader \citep{kolda2009tensor}, tensor products involved are significantly more complex compared to matrices. \\
Let be $\mathcal{X} (I_1 \times \dots \times I_n \times \dots \times I_N)$ and $\mathbf{U} (H \times I_n )$,
$$\mathcal{Y}=\mathcal{X} \times_n \mathbf{U} $$
with $\mathcal{Y} (I_1 \times \dots \times H \times \dots \times I_N)$, and where $\times_n $ is  the \textit{n}-mode product of a tensor.\\
The norm of a tensor $\mathcal{X}$  is analogous to the Frobenius norm and is defined as the square root of the sum of the squares of all its elements:
$$\left \|  \mathcal{X} \right \|= \sqrt{\sum_{i_1=1}^{I_1}\sum_{i_2=1}^{I_2}\dots\sum_{i_N=1}^{I_N}x^2_{i_1i_2\dots i_N}} $$
The inner product of two same-sized tensors $\mathcal{X}$,  $\mathcal{Y}$ is the sum of the products of their corresponding entries:
$$\left \langle \mathcal{X}, \ \mathcal{Y}  \right \rangle = \sum_{i_1=1}^{I_1}\sum_{i_2=1}^{I_2}\dots\sum_{i_N=1}^{I_N}x_{i_1i_2\dots i_N}y_{i_1i_2\dots i_N}. $$
The rank of a tensor is thoroughly different from its matrix counterpart.  Specifically, $\mathcal{X}$ has a rank-$(R_1, R_2, \dots ,R_N)$, known as multilinear rank, where $R_n$ is the dimension of vector space spanned by the \textit{n}-mode vectors:
$$R_n=\textrm{rank}_n(\mathcal{X})=\textrm{rank}(\mathbf{X}_{(n)})$$
with $R_n \leq I_n$ and $n =1 \dots N$.
	\subsection{Tucker model}
\label{sec:2.2}
In accord to the notation given in Section~\ref{sec:2.1}, the Tucker model for a \textit{N}th order tensor can be written as 
\begin{equation}\label{F2}
	\mathcal{X} \approx
	\mathcal{G} \times_1 \mathbf{U}^{(1)} \times_2 \mathbf{U}^{(2)}  \dots \times_n \mathbf{U}^{(n)}\dots \times_N  \mathbf{U}^{(N)}
\end{equation}
where each $\mathbf{U}^{(n)}  (I_n \times R_n)$ is an orthonormal component matrix ($\mathbf{U}^{(n)t}\mathbf{U}^{(n)}=\mathbf{I} $); and $\mathcal{G}$ is the core tensor with dimension $(R_1 \times R_2 \times  \dots \times R_N)$. \\
Assuming  $N=2$ with dimension $(I_1 \times I_2)$, and  $R_1$ and $R_2$ the dimension of vector space spanned by the first and second modes respectively, Tucker decomposition can be written as
\begin{equation}\label{F1}
	\mathcal{X} \approx
	\mathcal{G} \times_1 \mathbf{U}^{(1)} \times_2 \mathbf{U}^{(2)}
\end{equation}
where $\mathbf{U}^{(1)}  (I_1 \times R_1)$ and $\mathbf{U}^{(2)} (I_2 \times R_2)$ are orthonormal component matrices; and $\mathcal{G}$ is the core tensor with dimension $(R_1 \times R_2)$.\\
It is possible to show that $\mathbf{U}^{(1)}$ and $\mathbf{U}^{(2)}$ are the left and right singular vectors of $\mathcal{X}$, while the core tensor is a matrix with singular values on the first $r$ diagonal elements ($r=min(R_1,R_2)$) and 0s otherwise. \\
It is important underline that the core tensor for $N>2$ is a full tensor, where their elements can be negative too and their magnitude indicate the strength of interaction between components represented by the corresponding indices in the component matrices.

\subsection{Tucker parameters estimation}
\label{sec:2.3}
The Tucker decomposition of a \textit{2}nd order tensor is equivalent to a SVD. Thus, it is not a surprise that HOSVD is the most popular algorithm for estimate the parameters of a Tucker model in case of $N>2$.\\
Introduced by De Lathauwer et al. \citep{de2000multilinear}, which generalizes the three-ways case presented by Tucker \citep{Tucker:1966},  HOSVD finds the exact Tucker decomposition of rank-$(R_1,R_2, \dots ,R_N)$, and subsequently provides the core tensor $\mathcal{G}$. In essence, HOSVD initially computes the left singular vectors of $\mathbf{X}_{(n)}$ for $n =1 \dots N$. Finally, the transposes of these component matrices are multiplied with the original tensor $\mathcal{X}$ to obtain the core tensor.\\
When exact parameters are not required, as well as in case of compression, only the first $r_n$ columns of $\mathbf{U}^{(n)}$ are considerated, we need to solve the objective function 
\begin{equation}\label{F2a}
	\left \| 	\mathcal{X} -
	\mathcal{G} \times_1 \mathbf{U}^{(1)} \times_2 \mathbf{U}^{(2)}  \dots \times_n \mathbf{U}^{(n)}\dots \times_N  \mathbf{U}^{(N)} \right \|^2 
\end{equation} 
where $\mathbf{U}^{(n)}  (I_n \times r_n)$ is orthonormal; and $\mathcal{G}$ has dimension $(r_1 \times r_2 \times  \dots \times r_N)$. \\
The details of the optimization problem are readily available \citep{de2000best, kolda2009tensor}, when the version of HOSVD to estimate the parameters  is called \textit{truncated} HOSVD (\textit{t}-HOSVD). It is briefly described below highlighting its simplicity and limited number of steps. \\

{\small
	\begin{tabular}{l}\hline
		\textbf{Algorithm 1:} \textit{t}-HOSVD \\ \hline 
		\textbf{Require:} Tensor $\mathcal{X}\ (I_1,I_2, \dots, I_N)$ and ($r_1, \ r_2, \dots, \ r_N$)\\ 
		\textbf{for} {$n$ in $1,\dots,N$} \textbf{do}\\
		\ \ \ \ \ Consider the unfolding $\mathbf{X}_{(n)}$\\
		\ \ \ \ \  Compute the SVD($\mathbf{X}_{(n)}$)\\
		\ \ \ \ \  $\mathbf{U}^{(n)}\gets$ $r_n$ left singular vectors of $\mathbf{X}_{(n)}$\\
		\textbf{end for} \\
		$\mathcal{G} \gets \mathcal{X} \times_1 \mathbf{U}^{(1)t} \times_2 \mathbf{U}^{(2)t} \times_3 \dots \times_N \mathbf{U}^{(N)t}$\\
		return $\mathcal{G}$, $\mathbf{U}^{(1)}\dots \mathbf{U}^{(n)}\dots \mathbf{U}^{(N)}$  \\
		\textbf{end}	\\ \hline
		\label{alg1}
\end{tabular}}\\

Based on lossless compression strategy, HOSVD can reduce significantly the memory cost of the original tensor. Its effectiveness is determined by $\textrm{rank}-(R_1, R_2, \dots, R_N)$ compared the dimensions of modes $(I_1, I_2, \dots, I_N)$. In fact, original memory cost of $\mathcal{X}$ is $\prod_{n=1}^{N}I_n$, while the memory cost of HOSVD is given by a sum of smaller matrices and a core tensor, i.e $\sum_{n=1}^{N} I_n \times R_n$ elements of component matrices and $\prod_{n=1}^{N}R_n$ for core tensor.\\
Differently, by choose  $r_1 \ll R_1 \dots r_n \ll R_n \dots r_N \ll R_N$, \textit{t}-HOSVD assures an higher compression while still capturing the essential structure of the original tensor. However, the choice of parameters $(r_1, \ r_2, \dots, \ r_N)$ directly influences the balance between compression (memory savings) and accuracy (quality of the reconstruction of the original tensor).\\
Another crucial aspect to consider when algorithms are used for compression is their computational efficiency in terms of FLOPs (floatting points operation per seconds).
In the HOSVD algorithm, SVD is used to decompose $N$ flattened matrices $\mathbf{X}_{(n)}$, with $I_n \ll I$ ($I=\prod_{i \neq n} I_i$), thus computational cost for these steps is given by the sum of $N$ single SVDs:
\begin{equation}\label{F3}
	\sum_{n=1}^{N}(I_nI^2+\mathcal{O}(I^3)).
\end{equation}
	\section{\textit{ten}SVD for compression}
\label{sec:3}
In data compression context, fixed the accuracy of image or video, the goal is maximize the compression or saving memory space. In same way we introduce \textit{ten}SVD algorithm, but we want before pay attention on computational cost and its impact on energy and time consuming.\\
\subsection{Computational cost}
From Sections~\ref{sec:2.3}, it is easy verify that left singular vector of each $\mathbf{X}_{(n)}$ has $\mathcal{O}(I_nI^2+I^3)$ FLOPs. On the other side, it is clear that calculating the left singular vectors of $\mathbf{X}_{(n)}\mathbf{X}_{(n)}^t$ requires less computational overhead. 
In fact, we need $\mathcal{O}(II^2_n)$ FLOPs for multiple $\mathbf{X}_{(n)}$ with its transpose, and $\mathcal{O}(I_n^3)$ for the SVD of $\mathbf{X}_{(n)}\mathbf{X}_{(n)}^t$. Thus, the total cost for obtain left singular vectors is $\mathcal{O}(II^2_n+I_n^3)$. \\ 
Following this approach, the FLOPs required by HOSVD are
\begin{equation}\label{F4}
	\sum_{n=1}^{N}\mathcal{O}(I_nI^2+I^3_n).
\end{equation}
Due to $I_n \ll I$, $\mathcal{O}(I^3)$ is higher then $\mathcal{O}(I^3_n)$.\\
Based on this consideration, the primary concept of the new approach involves rearranging all elements of an \textit{N}th-order tensor $\mathcal{X}$, into an \textit{M}th-order tensor $\mathcal{Z}$ with $(M \ll N)$. Let be $(I_1, I_2, \dots, I_N)$ and $(J_1, J_2, \dots, J_M)$ the dimensionality of $\mathcal{X}$ and $\mathcal{Z}$ respectively. In this transformation, we have that $(I_n \geq J_m)$ for all $m$ and $n$, and ensuring that the total number of elements remains conserved through rearrangement, such that:
$$\prod_{n=1}^{N} I_n = \prod_{m=1}^{M} J_m.$$
Let be $J=\prod_{j \neq m} J_j$, the computational cost of $M$ matrices $\mathbf{Z}_{(m)}\mathbf{Z}_{(m)}^t$  is
\begin{equation}\label{F5}
	\sum_{m=1}^{M}\mathcal{O}(J_mJ^2+J^3_m).
\end{equation}
It is easy to observe that $J_mJ^2$ and $I_nI^2$ are the dominant terms in equation~\ref{F5} and ~\ref{F4}, which outweighs the cubic term $J_m^3$ and $I_n^3$ respectively. Hence, the reordering of the elements across more modes, reducing the computational cost from $\sum_{n=1}^{N}I_nI^2$ to $ \sum_{m=1}^{M}J_mJ^2$.\\

\subsection{Memory cost}
When lossy compression is the strategy, to reduce the memory cost the ($r_1, \ r_2, \dots, \ r_N$) are fixed lower than ($R_1, \ R_2, \dots, \ R_N$). And in case the elements are reordered into an higher order tensor, it is not clear if the memory cost will be higher. Storage requirement are given by the sum of memory cost for each component matrix $\sum_{n=1}^{N} I_n \times r_n$, and core tensor $\prod_{n=1}^{N} r_n$. If the elements are rearranged into an $M$th order tensor, memory storage required will be $\sum_{m=1}^{M}(J_m \times r_m)+ \prod_{m=1}^{M} r_m$. Thus it is clear that the dominant memory cost is due to the core tensor. \\
On the other side, it is possible to show that the energy (in terms of sum of the squared of elements) of a tensor is equal to the energy of core tensor, as \\

$	\left \| 	\mathcal{Z} -
\mathcal{G} \times_1 \mathbf{U}^{(1)} \times_2 \mathbf{U}^{(2)}  \dots \times_m \mathbf{U}^{(m)}\dots \times_M  \mathbf{U}^{(M)} \right \|^2$ \\

$= \left \| 	\mathcal{Z}  \right \|^2-2 \left< 	\mathcal{Z} \times_1 \mathbf{U}^{(1)t} \times_2 \mathbf{U}^{(2)t}  \dots \times_m \mathbf{U}^{(m)t}\dots \times_M  \mathbf{U}^{(M)t}, \mathcal{G} \right>+\left \| 	\mathcal{G}  \right \|^2$ \\

$= \left \| 	\mathcal{Z}  \right \|^2-2 \left< \mathcal{G},\mathcal{G} \right>+\left \| 	\mathcal{G}  \right \|^2 $ \\

$=\left \| 	\mathcal{Z}  \right \|^2-\left \| 	\mathcal{G}  \right \|^2.$\\

In Appendix A, it is shown too that core tensor energy is concentrated in a relatively small number of elements. Based on these two considerations, it could be reasonable consider and save only the elements of core that have the highest square values. Moreover, instead to save a fixed number of columns for each component matrices, i.e. $\mathbf{U}$ $(J_m \times r_m)$ for $m=1, \dots, M$, and a core tensor with correspondent dimension, it is possible to save the full component matrices, i.e. $\mathbf{U}$ $(J_m \times J_m)$ for $m=1, \dots, M$, and only the elements of core tensor with the highest magnitude, so that the ratio of energy collected is as high as researched.\\

\subsection{\textit{ten}SVD algorithm}
\begin{figure}[!ht]
	\centering
	\resizebox{1\textwidth}{!}{%
		\begin{circuitikz}
			\tikzstyle{every node}=[font=\LARGE]
			\draw  (1.25,12.75) rectangle (6,9);
			\draw [ fill={rgb,255:red,3; green,3; blue,3} , line width=0.2pt ] (17,6.75) rectangle (17.25,5);
			\draw  (7.25,12.75) rectangle (12,9);
			\draw  (13.25,12.75) rectangle (18,9);
			\draw [->, >=Stealth] (6,11) -- (7,11);
			\draw [->, >=Stealth] (12,11) -- (13,11);
			\draw  (1.25,7.5) rectangle (6,3.75);
			\draw [->, >=Stealth] (7.25,5.75) -- (6.25,5.75);
			\draw  (7.25,7.5) rectangle (12,3.75);
			\draw [->, >=Stealth] (13.25,5.75) -- (12.25,5.75);
			\draw  (13.25,7.5) rectangle (18,3.75);
			\draw [->, >=Stealth] (15.5,9) -- (15.5,7.5);
			\draw  (2.75,12.2) rectangle (4.35,10.2);
			\draw  (9,11.5) rectangle (10,10.25);
			\draw [short] (9,11.5) -- (9.5,12);
			\draw [dashed] (9,10.25) -- (9.5,10.75);
			\draw [, dashed] (9.5,12) rectangle  (10.5,10.75);
			\draw [short] (10,10.25) -- (10.5,10.75);
			\draw [short] (10,11.5) -- (10.5,12);
			\draw  (13.75,11.5) rectangle (14.75,10.25);
			\draw  (16.75,11.25) rectangle  node {\small $\mathbf{U}_2$} (17.75,10.5);
			\draw [short] (15.75,11.75) -- (16.25,12.25);
			\draw [short] (16.5,11.75) -- (17,12.25);
			\draw [short] (15.75,11.75) -- (16.5,11.75);
			\draw [short] (16.25,12.25) -- (17,12.25);
			\draw  (15.5,11) rectangle (16,10.5);
			\draw [short] (15.5,11) -- (15.75,11.25);
			\draw [short] (16,10.5) -- (16.25,10.75);
			\draw [short] (15.5,11) -- (15.75,11.25);
			\draw [, rotate around={-360:(16, 11)}, dashed] (15.75,11.25) rectangle  (16.25,10.75);
			\draw [dashed] (15.5,10.5) -- (15.75,10.75);
			\node [font=\Large] at (16,6) {$\Rightarrow $};
			\draw [, line width=0.2pt ] (14,6) rectangle (14.75,5.25);
			\draw [, line width=0.2pt , dashed] (14.5,6.5) rectangle  (15.25,5.75);
			\draw [line width=0.2pt, short] (14,6) -- (14.5,6.5);
			\draw [line width=0.2pt, short] (14.75,6) -- (15.25,6.5);
			\draw [line width=0.2pt, short] (14.75,5.25) -- (15.25,5.75);
			\draw [line width=0.2pt, dashed] (14,5.25) -- (14.5,5.75);
			\draw [, line width=0.2pt , dashed] (14.25,6) rectangle  (14.25,5.75);
			\node [font=\small] at (3.78,13.1) {\textbf{Phase 1:} Original tensor.};
			\node [font=\tiny] at (3.5,10.05) {$I_2$};
			\node [font=\tiny] at (2.6,11.3) {$I_1$};
			\node [font=\normalsize] at (9.6,11) {$\mathcal{Z}$};
			\node [font=\small] at (9.5,13.1) {\textbf{Phase 2:} Reshape tensor.};
			\node [font=\tiny] at (9.5,10.15) {$J_2$};
			\node [font=\tiny] at (8.9,11) {$J_1$};
			\node [font=\tiny] at (9.15,11.8) {$J_3$};
			\node [font=\tiny] at (13.65,11) {$J_1$};
			\node [font=\tiny] at (15.81,12.1) {$J_3$};
			\node [font=\tiny] at (17.22,10.26) {$J_2$};
			\node [font=\small] at (14.25,11) {$\mathbf{U}_1$};
			\node [font=\small] at (16.42,12) {$\mathbf{U}_3$};
			\node [font=\small] at (15.79,10.82) {$\mathcal{G}$};
			\node [font=\small] at (15.58,13.1) {\textbf{Phase 3:} Tucker decomposition.};
			\draw [, line width=2pt ] (14.5,5.5) rectangle (14.5,5.25);
			\draw [, line width=2pt ] (14.25,5.75) rectangle (14.25,5.5);
			\draw [, line width=2pt ] (14.5,6.25) rectangle (14.5,6);
			\draw [, line width=2pt ] (14.75,5.75) rectangle (14.75,5.5);
			\draw [, line width=2pt ] (15,6) rectangle (15,5.75);
			\draw [, line width=2pt ] (15,6.5) rectangle (15,6.25);
			\draw [, line width=2pt ] (14.75,6.5) rectangle (14.75,6.25);
			\node [font=\tiny] at (13.85,5.7) {$R_1$};
			\node [font=\tiny] at (14.3,5) {$R_2$};
			\node [font=\tiny] at (14.13,6.31) {$R_3$};
			\node [font=\tiny] at (15.45,11.23) {$R_3$};
			\node [font=\tiny] at (15.75,10.35) {$R_2$};
			\node [font=\tiny] at (15.26,10.75) {$R_1$};
			\draw [short] (16,11) -- (16.25,11.25);
			\node [font=\tiny] at (16.56,10.85) {$R_2$};
			\node [font=\tiny] at (16.22,11.6) {$R_3$};
			\node [font=\tiny] at (14.25,11.75) {$R_1$};
			\node [font=\small] at (15.6,3.3) {\textbf{Phase 4:} Larger absolute values.};
			\node [font=\small] at (9.5,3.3) {\textbf{Phase 5:} Store information.};
			\node [font=\small] at (17.15,4.7) {$\textbf{c}^t$};
			\draw [ fill={rgb,255:red,3; green,3; blue,3} , line width=0.2pt ] (8.5,4.75) rectangle (10.5,5);
			\node [font=\small] at (9.5,4.5) {\textbf{c}};
			\draw  (7.75,7.25) rectangle (8.75,6);
			\node [font=\small] at (8.25,6.75) {$\mathbf{U}_1$};
			\draw  (9.25,7.25) rectangle (10.25,6);
			\node [font=\small] at (9.75,6.75) {$\mathbf{U}_2$};
			\draw  (10.75,7.25) rectangle (11.75,6);
			\node [font=\small] at (11.25,6.75) {$\mathbf{U}_3$};
			\node [font=\normalsize] at (3.5,11.4) {$\mathcal{X}$};
			\node [font=\small] at (3.6,6.75) {$\mathcal{\widetilde{Z}}=\mathcal{G}\times_1\mathbf{U}_1\times_2\mathbf{U}_2\times_3\mathbf{U}_3$};
			\node [font=\small] at (3.5,5.5) {$\textrm{reshape}(\mathcal{\widetilde{Z}})\Rightarrow \mathcal{\widetilde{X}}$};
			\node [font=\small] at (3.8,3.3) {\textbf{Phase 6:} Reconstructed tensor.};
	\end{circuitikz}}%
	\caption{\textit{ten}SVD framework.}
	\label{fig:1}
\end{figure}
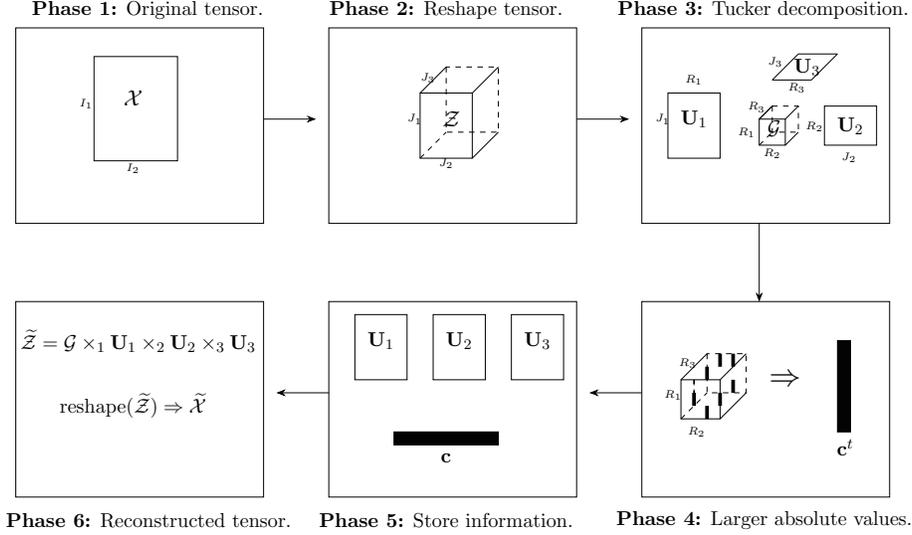

To show the main phases of the \textit{ten}SVD procedure, in Figure~\ref{fig:1} a matrix (2nd order tensor) is reshaped into a 3rd order ones.\\
According to several strategies, such as compression ratio or the percentage of original information to retain, the quality measure of the original object is fixed (Phase 1). 
In Phase 2, the order and dimensions of the new object must be decided, where the order is higher than that of the original tensor and the dimensions for each mode are lower. Heuristically, we have observed that \textit{ten}SVD is more efficient when the new tensor is close to a hypercube. 
In Phase 3, component matrices and the core tensor are calculated using SVD. 
Elements of the core tensor with highest magnitude are selected to ensure the desired level of quality (Phase 4). 
In Phase 5, component matrices and selected elements with their relative positions in the core tensor are saved in $\mathbf{U}$ and $\mathbf{c}$, respectively. Data saved in this phase could be stored or transmitted to save space (in case of backup) and bandwidth (in case of transmission).
In the last Phase, the saved information is used to reconstruct the approximated original image.\\
The \textit{ten}SVD steps are detailed given in Algorithm 2, where it is possible to observe that both accuracy, given by Relative Error ($\epsilon$), or compression, given by compression ratio (com), can be fixed. \\
In first stage of the procedure, to rearranged the elements into an $M$-order tensor with dimensions $J_1, J_2, \dots, J_M$, total number of elements, i.e. $\prod_{i=1}^{N}I_i$, is
divides in prime factors, ordered increasing way smaller values are aggregated until the minimum of them is about equal to $M$.\\
In second stage, HOSVD algorithm is applied on new tensor $\mathcal{Z}$. Full left singular vectors and core tensor is given. From computational point of view, it is more efficient to work on $\mathbf{Z}_{(m)}\mathbf{Z}_{(m)}^t$ then on the $m$-mode unfolding $\mathbf{Z}_{(m)}$.\\
In the last stage, the highest absolute values and their position in core tensor are selected in a vector as well as the relative error or level of compression fixed is assured. Finally, these elements and component matrices are saved. 
{\small
	\begin{tabular}{l}\hline
		\textbf{Algorithm 2:} \textit{ten}SVD \\ \hline 
		\textbf{Require:} Tensor $\mathcal{X} (I_1,I_2, \dots, I_N)$; $\epsilon$ (or com)\\ 
		Reshape  $\mathcal{X} \Rightarrow  \mathcal{Z} (J_1,J_2, \dots, J_J)$,  \\ 	\ \ \ \ \  with $J_1 \cong J_2 \cong \dots \cong J_M \cong M$ and $\prod_{i=1}^{N}I_i=\prod_{m=1}^{M}J_m$ \\
		\textbf{for} {$m$ in $1,\dots,M$} \textbf{do}\\
		\ \ \ \ \  Consider the unfolding $\mathbf{Z}_{(m)}$ of $\mathcal{Z}$\\
		\ \ \ \ \  $\mathbf{U}^{(m)}\gets$ $R_m$ left singular vectors of SVD($\mathbf{Z}_{(m)}\mathbf{Z}_{(m)}^t$)\\
		\textbf{end for} \\
		$dim\gets size \left (\mathbf{U}^{(1)} \right)+\dots + size \left(\mathbf{U}^{(M)} \right)$\\
		$\mathcal{G} \gets \mathcal{Z} \times_1 \mathbf{U}^{(1)t} \times_2 \mathbf{U}^{(2)t} \times_3 \dots \times_M \mathbf{U}^{(M)t}$\\
		Select element of $\mathcal{G}$ \\
		$\mathbf{g}\gets vec(abs(\mathcal{G})) $ \\
		Order in decreasing way the  elements of $\mathbf{g}$ \\
		$r \gets 1$ \\
		$\mathbf{\hat{g}} \gets 0$ \\
		\textbf{while} {$\left \| sum(\mathbf{g}) - sum(\mathbf{\hat{g}}) \right \|/ \left \| sum(\mathbf{g}) \right \| \geq \epsilon$} (or $dim /size \left( \mathcal{Z}\right) \leq com$) \textbf{do}\\
		\ \ \ \ \ $\mathbf{\hat{g}} \gets \mathbf{\hat{g}} + \mathbf{g}[r]$ \\
		\ \ \ \ \ $r \gets 1 + 1$ \\
		\ \ \ \ \  Set in $\mathbf{c}$ the elements of $\mathcal{G}$ selected and their relative position\\
		\ \ \ \ \  $dim \gets dim+size(\mathbf{c})$	\\ 		
		\textbf{end while} \\
		return $\mathbf{c}$, $\mathbf{U}^{(1)}\dots \mathbf{U}^{(M)}$,  $I_1 \dots I_N$ \\
		\textbf{end}	\\ \hline
		\label{tab:4}
\end{tabular}}\\
		\section{Case study}
\label{sec:4}
\begin{figure}[htbp]
			\centering
\includegraphics[bb=0 -100 640 480, scale=0.5]{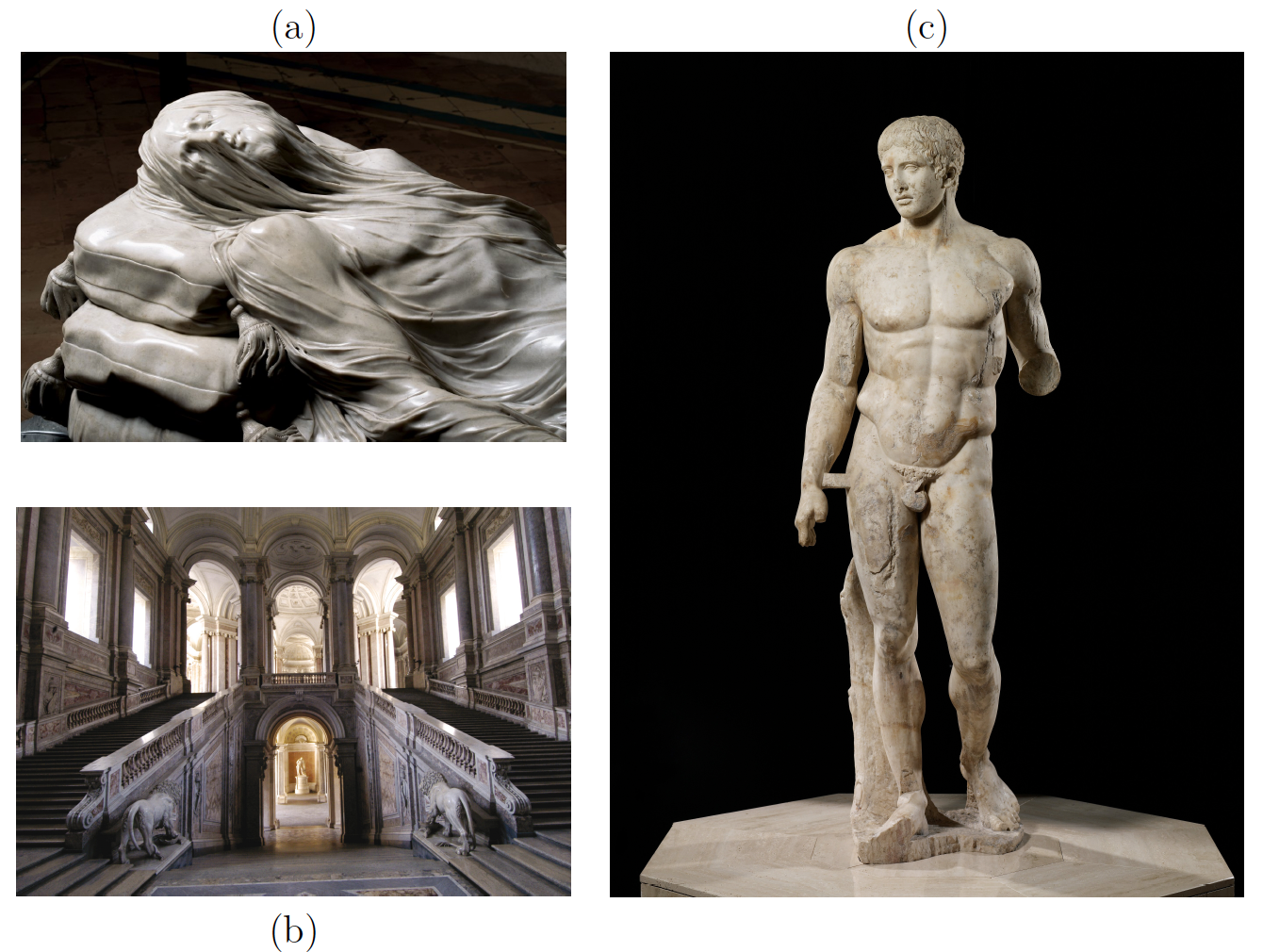} \vspace{-2cm}
	\caption{The Veilled Christ (a),the Royal Palace of Caserta (b), the Dorophos (c)}
	\label{fig:4}
\end{figure}

\textit{ten}SVD algorithm was  implemented in R and, to evaluate the performance for lossy compression problem, it is compared with HOSVD algorithm (package ‘tensr’ and 'rrcov3way'). The analysis was conducted using an Intel(R) Xeon(R) Gold 6288 CPU 2.10GHz 2.10 GHz (16 cores) with 512GB RAM, and R version 4.2.0.\\
Three pictures representing the Veiled Christ, the Dorophos, and the Royal Palace of Caserta are shown in Fig.~\ref{fig:4}. They are converted in datasets, Fig.~\ref{fig:4}~(a) has 880 rows and 1,240 columns, Fig.~\ref{fig:4}~(b) has 1,792 rows and 2,560 columns, and Fig.~\ref{fig:4}~(c) has 7,200 rows and 5,400 columns, all with three channels (RGB). Thus, three third-order tensors with dimensions $(880 \times 1,240 \times 3)$, $1792\times2560\times3$, and $7200\times5400\times3$ were generated. Moreover, a short video of the Athletics 100 meters final of the Olympic Games Paris 2024 has been acquired. Package grDevices 4.4.0 was used to convert the three pictures in tensors. Av package 0.9 was used to split the video in png images, where the rate was fixed in three images per second. These frames were converted in a fourth tensor with dimension $(30 \times 360 \times 640 \times 3)$.

		\subsection{Evaluation Metrics}\label{sec:4.1}
Mean Squared Error (MSE), Relative Error (ERR ) and Peak-Signal-to-Noise Ratio (PSNR) are widely used to measure the accuracy of Tucker decomposition, while PSNR is used to measure the quality of compressed images or videos. And they are defined as follows\\

\begin{itemize}
	\item MSE is the square of the Frobenius norm of the difference between predicted and observed values of a tensor, normalized by the number of tensor elements: 
	$$ \textrm{MSE}=\frac{\left \| \mathcal{X}-\mathcal{\widehat{X}} \right \|^2}{I_1*I_2*\dots*I_N}$$
	\item ERR quantifies the difference between the original tensor and its approximation, normalized by the magnitude of the original tensor
	$$ \textrm{ERR}=\frac{\left \| \mathcal{X}-\mathcal{\widehat{X}} \right \|}{\left \| \mathcal{X}\right \|}$$
	\item PSNR computes the peak signal-to-noise ratio between two images in decibels (dB). This ratio is a quality measurement between original and compressed image. PSNR can take values up to infinity; the higher the PSNR, the better the compressed image quality: 
	$$ \textrm{PSNR}=10\textrm{log}_{10}\left (\frac{\textrm{max}(\textrm{max}(\mathcal {X}),\textrm{max}(\mathcal {\widehat{X}}))^2}{\sqrt{\textrm{MSE}}}  \right )$$
\end{itemize}

\begin{figure}[htbp]\vspace{-3cm}
		\centering
\includegraphics[bb=0 0 640 480, scale=0.5]{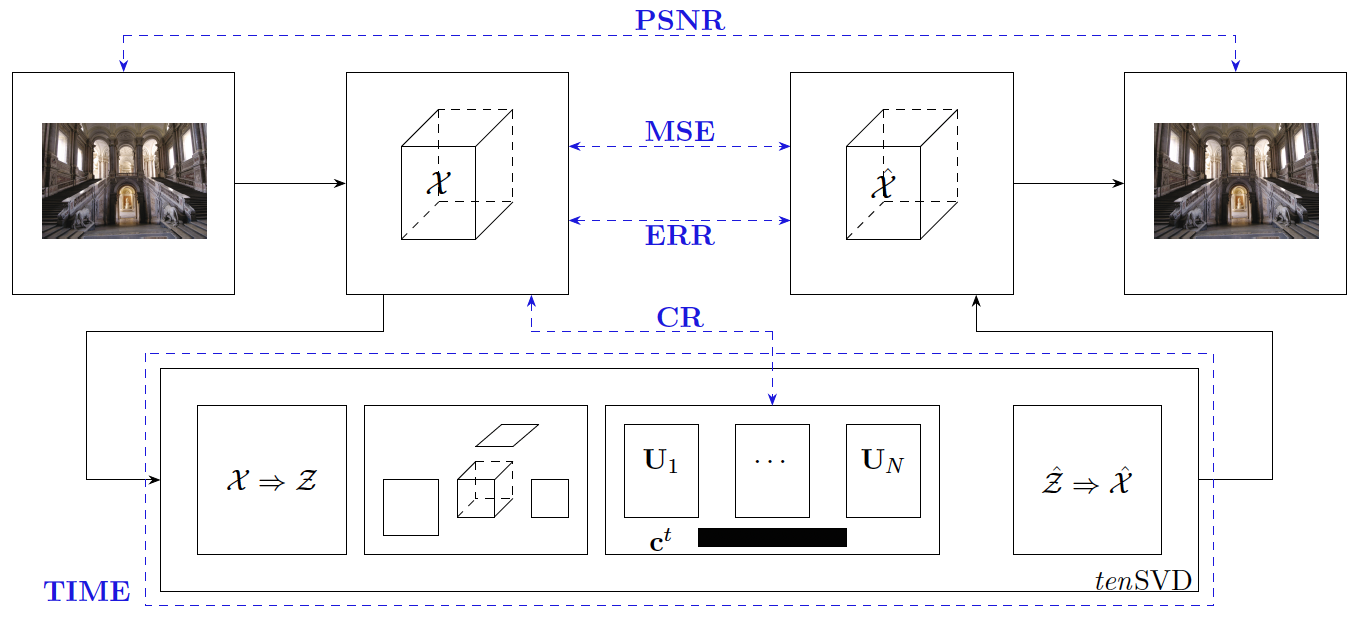}
	\caption{Used measurement protocol}
	\label{fig:4.2}
\end{figure}
Moreover, to measure compression effectiveness in term of space and time, two other metrics are used:

\begin{itemize}
			\item Compression Ratio (CR) is the ratio between the size of the original tensor and the number of elements of the compression representation (storage cost).
	$$ \textrm{CR}=\frac{\textrm{compressed \ data \ size}}{\textrm{original \ data \ size}}$$
	\item Processing time (TIME) is the real elapsed time since the algorithm was started.
\end{itemize}
The detailed scheme of the developed framework is shown in Figure~\ref{fig:4.2}. 

		\subsection{Results}
\label{sec:4.2}
\begin{figure}[htbp]\vspace{-3cm}
		\centering
\includegraphics[bb=0 0 640 480, scale=0.5]{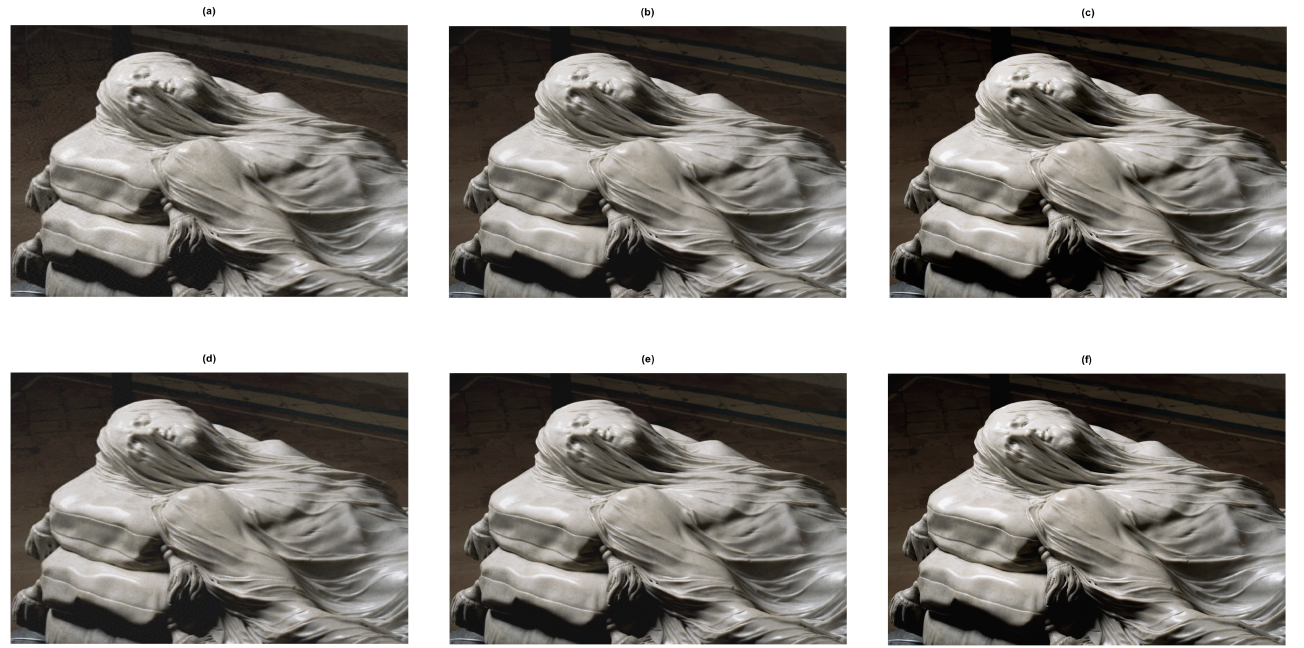}
	\caption{Algorithm HOSVD CR$=93.7 \%$ (a), CR$=83.4 \%$ (b), and CR$=28.2 \%$ (c); Algorithm \textit{ten}SVD CR$=93.7 \%$ (d), CR$=83.4 \%$ (e), and CR$=28.2 \%$ (f).}
	\label{fig:5.1}
\end{figure}

\begin{figure}[htbp]\vspace{-3cm}
		\centering
\includegraphics[bb=0 0 640 480, scale=0.5]{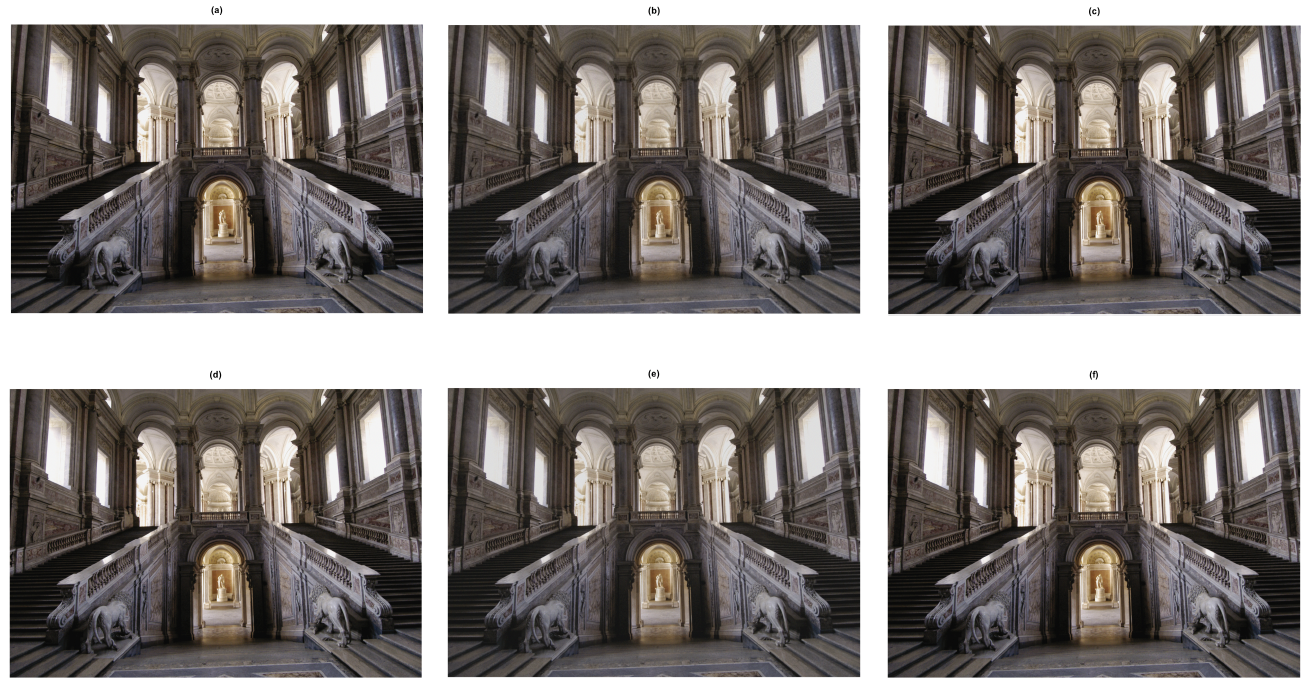}
	\caption{Algorithm HOSVD CR$=96.6\%$ (a), CR$=92.8 \%$ (b), and CR$=83.2 \%$ (c); Algorithm \textit{ten}SVD CR$=96.6 \%$ (d), CR$=92.8 \%$ (e), and CR$=83.2 \%$ (f).}
	\label{fig:5.2}
\end{figure}

\begin{figure}[htbp]
	\centering
 \includegraphics[bb=0 0 640 480, scale=0.5]{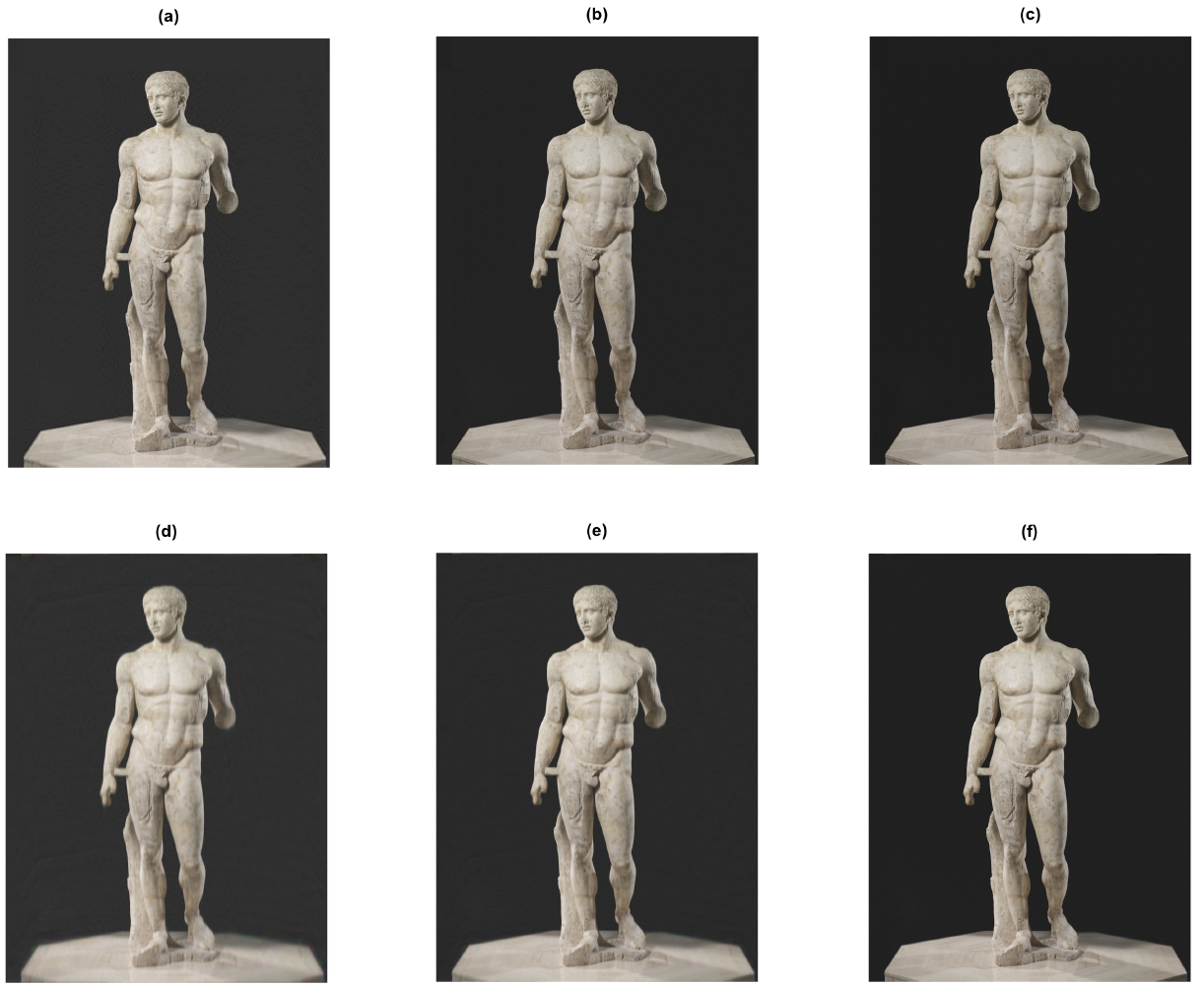}
	\caption{Algorithm HOSVD CR$=98.9 \%$ (a), CR$=97.7 \%$ (b), and CR$=92.5 \%$ (c); Algorithm \textit{ten}SVD CR$=98.9 \%$ (d), CR$=97.7 \%$ (e), and CR$=92.5 \%$ (f).}
		\label{fig:5.3}
\end{figure}

To compress the images, three different sets of parameters are chosen for HOSVD, $r_1=100$, $ r_2=100$, $ r_3=3$ (a),  $r_1=200$, $ r_2=200$, $ r_3=3$ (b),  $r_1=600$, $ r_2=600$, $ r_3=3$ (c), and CR values observed by \textit{t-}HOSVD are used as parameters into \textit{ten}SVD in order to get comparable results in term of memory cost.\\ 
In Figures~\ref{fig:5.1},~\ref{fig:5.2}, and~\ref{fig:5.3} are shown the reconstructed images with HOSVD (upper side) and \textit{ten}SVD (bottom side). The different scenarios show that both algorithms ensure the same accuracy in term of quality.\\

\begin{figure}[htbp]
	\centering
	\includegraphics[bb=0 0 640 480, scale=0.5]{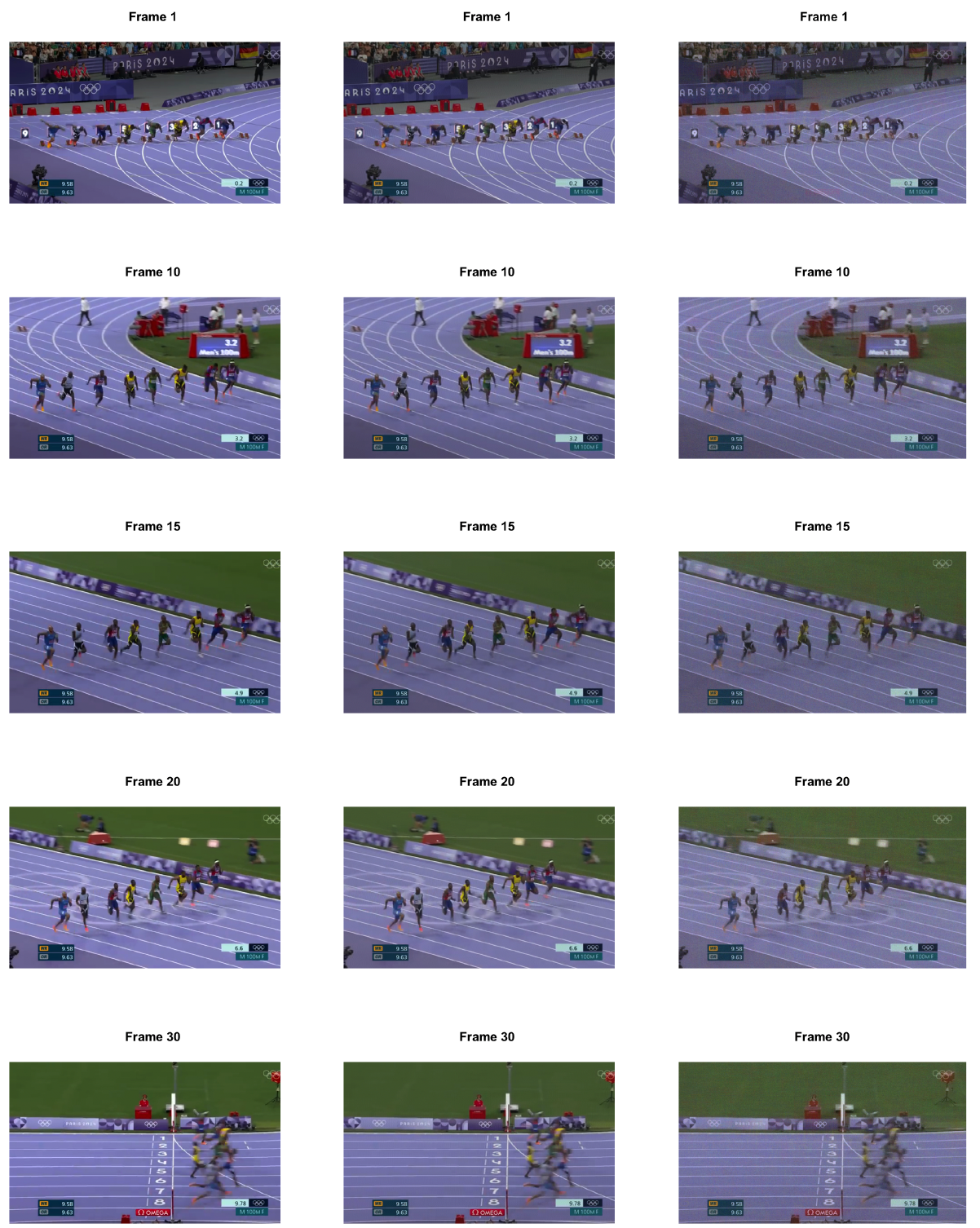}
\caption{Frames 1, 10, 15, 20, 30, original on left side, HOSVD on central and \textit{ten}SVD on right side}
\label{fig:6ee}
\end{figure}

In Table~\ref{tab:compression} all diagnostic measures are summarized. It is clear that all scenarios show very comparable values in term of accuracy and quality as well as showed into Figures~\ref{fig:5.1}~\ref{fig:5.2}~\ref{fig:5.3}, but with fixed compression ratios, it is possible to observe that \textit{ten}SVD assure always best results in term of efficiency. It is four 4 times faster than \textit{t-}HOSVD for the Veilled Christ and 26 times faster for the Dorophos, both results is given for scenario (c).
\begin{table}[h!]
	\centering
	\renewcommand{\arraystretch}{0.2}
	\setlength{\tabcolsep}{6pt}
	\begin{tabular}{@{}llcrccccc@{}}
		\toprule
		\multirow{6}{*}{\textbf{Image}} & 	\multirow{6}{*}{\textbf{}} & 	\multirow{6}{*}{\textbf{Alg.}}   & 
		\multicolumn{5}{c}{\textbf{Metrics}} \\
		\cmidrule(lr){4-8} 
		& & & \textbf{CR} & \textbf{Time} & \textbf{PSNR} &\textbf{MSE}& \textbf{ERR}\\
		\midrule
		& \multirow{2}{*}{(a)} &H & $93.7 \%$ & 28  & 24.2 &0.0037&0.147\\
		&  &T & $93.7 \%$& 5  & 23.8 &0.0041& 0.157\\ 			
		\rule{0pt}{5pt} \\
		\multirow{6}{*}{Figure 4}& \multirow{2}{*}{(b)} &H   & $83.4 \%$& 29  & 27.5 &0.0017& 0.103\\
		&  &T  & $83.4 \%$&5  & 27.o &0.0019& 0.108\\				
		\rule{0pt}{5pt} \\
		& \multirow{2}{*}{(c)} &H   &$28.2 \%$& 33  & 38.9 &0.0001& 0.027\\
		&  &T  & $28.2 \%$& 8 & 34.9 &0.0003& 0.043\\
		\midrule
		& \multirow{2}{*}{(a)} &H & $96.6 \%$ & 271  & 25.3 &0.0029&0.107\\
		&  &T  & $96.6 \%$& 24 & 24.5 &0.0034& 0.118\\	
		\rule{0pt}{5pt} \\
		\multirow{6}{*}{Figure 5 }& \multirow{2}{*}{(b)} &H   & $92.8 \%$& 273  & 26.2 &0.0023& 0.099\\
		& & T  & $92.8 \%$&27  & 26.0 &0.0024& 0.101\\				
		\rule{0pt}{5pt} \\
		& \multirow{2}{*}{(c)}& H   &$83.2\%$& 286  & 28.2 &0.0015& 0.101\\
		&  &T  & $83.2\%$& 31 & 29.5 &0.0011& 0.087\\
		\midrule
		& \multirow{2}{*}{(a)} &H & $98.9 \%$ & 6681  & 16.0 &0.0252&0.94\\
		&  &T  & $98.9 \%$& 248 & 15.5 &0.0281& 0.99\\	
		\rule{0pt}{5pt} \\
		\multirow{6}{*}{Figure 6 }& \multirow{2}{*}{(b)} &H   & $97.7 \%$& 6703  & 18.4 &0.0143& 0.685\\
		&  & T  & $97.7 \%$&240  & 18.6 &0.021& 0.855\\				
		\rule{0pt}{5pt} \\
		& \multirow{2}{*}{(c)} &H   &$92.5\%$& 6859  & 20.0 &0.0099& 0.57\\
		& & T  & $92.5\%$& 257 & 19.1 &0.0128& 0.64\\
		\bottomrule
	\end{tabular}
	\caption{Performance comparison of HOSVD (H) and \textit{ten}SVD (T) for Veilled Christ, Royal Palace of Caserta and Dorophos with different compression ratio scenarios.}
	\label{tab:compression}
\end{table}
Regarding the Olympic Games Paris 2024 videos, a fixed set of parameters was analyzed: $r_1=30, \ r_2=300, \  r_3=300, \ r_4=3$. The HOSVD yielded the following metrics:  MSE$=0.0013$, ERR$= 0.0594$, PSNR$=28$, CR$=41 \%$. In contrast, \textit{ten}SVD analyzed the same compression ratio, producing comparable metrics:  MSE$=0.0036$, ERR$= 0.0952$, PSNR$=24$The \textit{ten}SVD algorithm was approximately 2.5 times faster than HOSVD (22.024 seconds vs. 54.403 seconds). In Figure~\ref{fig:6ee} presents the original frames 1, 10, 15, 20, and 30 (on the left side), the reconstructed frames after HOSVD (in the center), and the reconstructed frames after applying \textit{ten}SVD (on the right side).

\section{Conclusion and findings}
\label{sec:5}
\begin{figure}
	\centering
	\includegraphics[bb=20 0 400 124,scale=0.85]{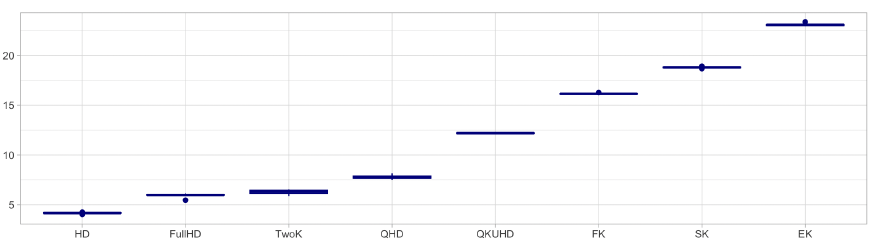}
	\caption{Computational time ratio (HOSVD/\textit{ten}SVD) across different image standards}
	\label{fig:7ff}
\end{figure}

For the same level of quality and compression, it is shown that \textit{ten}SVD is a faster algorithm than HOSVD. Moreover, it was possible to observe that results could be impressively faster in several scenarios. In Table \ref{tab:compression} it is possible to observe that in average \textit{ten}SVD is about 5.8 times faster than HOSVD for the 3,273,600 elements of $880 \times 1,240 \times 3$ tensor, 10.1 times for the 13,762,560 elements of $1,792 \times 2,560 \times 3$ tensor, and 27.9 times for the 32,531,040 elements of $2,378 \times 4,560 \times 3$ tensor. When it is only 2.5 times faster for the 20,736,000 elements $30 \times 360 \times 640 \times 3$ tensor. Thus, the difference of computational time depend from the shape more than size of tensor.\\   
To verify this, a simulation study was run on 8 different image standards: $1280 \times 720\times 3$(HD), $1920\times 1080 \times 3$ (FullHD), $2048\times1080\times3$ (TwoK), $2560\times1440\times3$ (FourHD), $3840\times2160\times3$ (FourKUHD), $5120\times2880\times3$ (FiveK), $6144\times3456\times3$ (SixK), $7680\times4320\times3$ (EightK). Datasets were generated 25 times by uniform distribution. Time was measured in seconds and summary statistics are given in Table~\ref{tab:compression}, when in Figure~\ref{fig:7ff} shows the elapsed time ratio between the two algorithms. \textit{ten}SVD is always faster than HOSVD  and the best results is for EightK scenario.\\ 

Our heuristic, it seems that \textit{ten}SVD is higher efficient when it can reshape original tensor in close to hypercube. Other elements that influence the efficiency is starting shape of tensor. In case of higher quality images, tensor is necessary high unbalanced, because one mode has dimension 3 (RGB channels). In these cases \textit{ten}SVD gives a huge performance in term of computational cost.\\    
Unfortunately, \textit{ten}SVD is not suitable in several other applications. It cannot find the true latent variables that generate data, because original structure is missed when tensor is reshaped in an higher-order ones. If the total of core energy doesn't change when tensor is reshaped, it is not possible to get the true orthonormal component matrices. In this contest, HOSVD works but it accuracy could be low in presence of significant noise. Iterative algorithms as Higher-Order Orthogonal Iteration (HOOI) \citep{kroonenberg1980, de2000best}, Sequentially Truncated Higher-Order Singular Value Decomposition (\textit{st}HOSVD) \citep{fang2018sequentially} or Gradient Descent methods generally yields higher accuracy \citep{10.1007/978-3-031-18907-4_32} \citep{trendafilov:mda}. 
These algorithms have advantage to converge at better fit and generally they are versatile and can be adapted to many different scenarios and constraints. On the other side, their iterative nature might get stuck in local minima and could be very expensive in term of time consuming and relative sustainability in term of energy required.\\

\bibliographystyle{apalike}  
\bibliography{mybibliography.bib}   

\end{document}